\documentclass[aps,preprint,prl]{revtex4}

\usepackage{amsmath}
\usepackage{amssymb}
\usepackage{graphicx}
\usepackage{bm}

\begin{document}

\title{Ultimate charge sensitivity and efficiency of a quantum point contact with a superposed input state}%
\author{Kang-Ho Lee}
\author{Kicheon Kang}
\email{kicheon.kang@gmail.com}
\affiliation{Department of Physics, Chonnam National University,
 Gwangju 500-757, Korea}
\date{\today}

\begin{abstract}
We address the ultimate charge detection scheme with a quantum point contact.
It is shown that a {\it superposed
input state} is necessary to exploit the full sensitivity of a quantum point contact detector.
The coherence of the input state provides an improvement in charge sensitivity, and
this improvement is a result of the fundamental property of the scattering matrix. Further, a quantum-limited (maximally efficient) detection is possible by controlling the interference between the two output waves. Our scheme provides the ultimate sensitivity and efficiency of charge detection with a generic quantum point contact.
\end{abstract}


\maketitle
\newcommand \tr {{\rm Tr}}

Detection of single electrons~\cite{Field93,Devoret00,Lu,Sprinzak02} is an essential ingredient for realizing quantum information processing
with charge qubits.
A quantum point contact (QPC) is widely used as a charge detector, with a sensitivity that extends down to the level of single electrons. It also plays an important role in investigating fundamental issues in quantum theory, such as
quantum mechanical complementarity~\cite{Buks98,Chang08}.
It has been well understood that the phase, as well as the transmission probability, of a
QPC can be utilized for charge sensing~\cite{Stodolsky99,Sprinzak00}. The sensitivity can be described in
terms of the controlled dephasing rate of the qubit induced by the interaction with the QPC detector. This is because the dephasing rate is equivalent to the rate of the (qubit) information transfer
to the detector.
The dephasing rate is a function of the two independent variables,
$\Delta T$ and $\Delta\phi$, the sensitivities of the transmission
probability and of the phase difference between the transmitted and the
reflected waves,
respectively~\cite{Aleiner97,Levinson97,Gurvitz97,Stodolsky99,Hackenbroich01}.
Further, it has been shown that the phase-sensitive term
is dominant in a generic QPC which reduces considerably the efficiency of
charge detection (typically, below 5$\%$)~\cite{Chang08,Kang05,Kang07}.
In this context, utilizing the
phase degree of freedom is important and useful for a quantum information
architecture.

On the other hand, it is worth considering the following unnoticed, but fundamental,
property of a QPC in the context of charge sensing.
A single channel QPC is described by a $2\times2$ scattering matrix, which has SU(2) symmetry.
Neglecting the physically irrelevant global phase factor, the S-matrix has
three independent physical variables for charge detection. However, all the existing
experiments and theoretical proposals are based on utilizing only
one or two
($\Delta T$ and $\Delta\phi$) of
these variables~\cite{Aleiner97,Buks98,Chang08,Sprinzak00,Levinson97,Gurvitz97,Stodolsky99,Hackenbroich01}.
Therefore, this provides an interesting question: can we exploit the three independent variables for
charge detection?

In this Letter, we show that this is indeed possible.
In addition to the well-known
sensitivities $\Delta T$ and $\Delta\phi$, another hidden
phase variable exists, and can indeed be used. We propose a scheme that utilizes this hidden variable
(as well as the two other variables) by using a ``superposed input state".
The hidden phase variable appears in the expression of the
dephasing rate, if the input electron is in the state of a coherent
superposition of two input ports.
Naturally, it provides an advantage in high-sensitivity charge detection
as well as deeper understanding of the quantum mechanical complementarity
realized in a QPC detector. The setup proposed here provides the maximum
sensitivity with a generic QPC. Further, we show that the system can be
tuned for a quantum limited detection of the charge state.
%

One of the most remarkable features in quantum measurement is
the trade-off between information transfer of the state of the {\em system}
into the measurement {\em apparatus}
and the back-action dephasing of the system~\cite{Korotkov01,Pilgram,Clerk,Averin05}.  The
``potential" measurement sensitivity of a measurement apparatus is
reflected in the dephasing rate induced by the apparatus.
 For an actual measurement, the information stored in the potential
sensitivity should be
transformed to an actual sensitivity. In general, the potential sensitivity
may not be fully exploited in
an actual measurement. The quantum-limited detection is a fully efficient measurement
where the potential sensitivity is fully transformed to the actual
sensitivity.
For a practical quantum information processing, both the sensitivity and
the efficiency are important:
Dephasing rate (sensitivity) is the speed of the information transfer,
and the efficiency is the ratio of the actual measurement rate to the dephasing
rate~\cite{Korotkov01,Pilgram,Averin05,Khym06}.

Let us consider a QPC charge detector, which monitors the state
of a charge qubit (being 0 or 1) through mutual capacitive
interactions~(Fig.~1).
Controlled dephasing induced by a charge detection can be implemented,
for instance, by constructing interferometers which include a
quantum dot~\cite{Buks98,Chang08} or double quantum dots~\cite{Sprinzak00}.
We assume that
the QPC circuit has only a single transverse channel at zero
temperature. Generalization to finite temperature and multichannel is
straightforward.  The interaction between the qubit and the QPC detector
is described as a continuous weak measurement~\cite{Korotkov01,Hackenbroich01}.
The sensitivity of a possible measurement is encoded in the scattering matrix of the QPC, which
depends on the qubit state $j$ ($=0$ or $1$):
\begin{equation}
{S_j}=
\left(\begin{array}{ccc}
r_j & t'_j \\
t_j & r'_j \\
\end{array} \right)~.
\label{eq:Sj}
\end{equation}
The scattering matrix transforms the input states $\alpha$ and $\beta$ into the
output $\gamma$ and $\delta$ as
\begin{equation}
 \left(\begin{array}{c}
  c_\gamma \\ c_\delta
 \end{array} \right)
= S_j
 \left(\begin{array}{c}
  c_\alpha \\ c_\beta
 \end{array} \right) ,
\end{equation}
where $c_l$ is the annihilation operator of an electron at lead $l (\in \alpha,\beta,\gamma,\delta)$.
%
For a single scattering event in the QPC detector,
the initial state of the system before scattering can be represented as a
product state of the two subsystems:
\begin{equation}
 |\Psi_0\rangle=(a_{0}|0\rangle+a_{1}|1\rangle)\otimes|\chi_{in}\rangle ,
\end{equation}
where $a_{0}|0\rangle+a_{1}|1\rangle$ is the initial state of the charge
qubit, and $|\chi_{in}\rangle$ is the input state of the QPC detector.

Our strategy here is to introduce a {\em superposed input} state from the two input
sources $\alpha$ and $\beta$:
\begin{equation}
 |\chi_{in}\rangle =
 (\sqrt{p}c^{\dagger}_\alpha+\sqrt{1-p}e^{i\theta}c^{\dagger}_\beta)|F\rangle,
\label{eq:input}
\end{equation}
instead of the conventional way of injecting the probe electrons from a
single source.
The parameters $p$ and $\theta$ determine the degree of splitting and
the relative phase between the two input waves, respectively. In a real experiment,
these parameters can be tuned by placing another QPC, before injecting electrons
into the region of the interactions. $|F\rangle$ denotes the ground state (Fermi sea)
of the electrodes.

Upon a scattering, the system is entangled as
\begin{equation}
 |\Psi\rangle = a_{0}|{0}\rangle|\chi_{0}\rangle
              + a_{1}|{1}\rangle|\chi_{1}\rangle ,
\end{equation}
where the output state of the QPC detector $|\chi_{j}\rangle$ is given by
\begin{subequations}
\begin{eqnarray}
 |\chi_{j}\rangle &=& (\tilde{r_j}c_{\gamma}^{\dagger}+\tilde{t_j}c_{\delta}^{\dagger})|{F}\rangle~,
   \label{eq:output} \\
 \tilde{r_j} &=& \sqrt{p}r_j+\sqrt{1-p}e^{i\theta}t'_j~, \\
 \tilde{t_j} &=& \sqrt{p}t_j+\sqrt{1-p}e^{i\theta}r'_j~.
\end{eqnarray}
\end{subequations}
Charge sensitivity is reflected in the reduced density
matrix of the qubit, $\rho=\rm{Tr}_{QPC}$\{$|\Psi\rangle\langle\Psi|$\}.
Upon a single scattering event, its off-diagonal element $\rho_{01}$ is reduced ($\rho_{01}\rightarrow
\lambda\rho_{01}$) by
the coherence factor $\lambda$
\begin{equation}
 \lambda = \langle\chi_1|\chi_0\rangle .
\label{eq:lambda}
\end{equation}
%

We consider the continuous weak measurement limit, where the single scattering
event provides only a slight modification of the qubit state ($\lambda\approx1$). The
scattering through the QPC takes place on a time scale much shorter than
the relevant time scale in the qubit. In our particular case of a QPC with
the applied bias voltage $V$, this corresponds to
$\Delta{t}\ll 1/\Gamma_d$, where $\Delta{t}\equiv h/eV$ is the average
time interval~\cite{Delta_t}  between two successive scattering events, and
$\Gamma_d$ is the dephasing rate.
In this process, the magnitude of $\rho_{01}$ decays as
\begin{equation}
 |\rho_{01}|=e^{-\Gamma_dt}|\rho^{0}_{01}|
\end{equation}
with the dephasing rate
$\Gamma_d=-\frac{\ln{|\lambda|}}{\Delta{t}}$.
In a conventional scheme with single input port ($p=0$ or $p=1$), the dephasing
rate is determined by the charge sensitivities of the two independent
parameters, namely $T_j=|t_j|^2$ and $\phi_j = \arg{(t_j/r_j)}$.
This is because the qubit state
information can be extracted either through the transmission
probability (with a direct current measurement), or through the relative phase shift between the transmitted
and  the reflected output waves (by constructing an interferometer). On the other hand, our scheme provides an
additional sensitivity on the parameter $\varphi_j\equiv \arg{(t_j/r_j')}$,
and the dephasing rate is given as
\begin{eqnarray}
 \Gamma_d &=& \frac{1}{\Delta t} \left[
     u_1(\Delta T)^2 + u_2(\Delta\phi)^2 + u_3(\Delta\varphi)^2
                                 \right. \nonumber \\
   &+& \left. u_4(\Delta T\Delta\phi) + u_5(\Delta T\Delta\varphi)
           +  u_6(\Delta\phi\Delta\varphi)
       \right] ,
\label{eq:Gamma_d}
\end{eqnarray}
with parameter-dependent dimensionless coefficients
$u_i$ ($i=1,2,\cdots,6$). $\Delta T$ is the sensitivity of the transmission
probability, that is, $\Delta T=|t_1|^2-|t_0|^2$. The phase sensitivities
are defined in the same way as $\Delta\phi=\phi_1-\phi_0$ and
$\Delta\varphi=\varphi_1-\varphi_0$.

The key point of Eq.~(\ref{eq:Gamma_d}) is that $\Gamma_d$ is
a function of the three independent charge sensitivities, $\Delta T$,
$\Delta\phi$, and $\Delta\varphi$, in contrast to
the well-known expression of the dephasing rate having only two
sensitivities, $\Delta T$ and
$\Delta\phi$~\cite{Aleiner97,Stodolsky99,Hackenbroich01}. The physical meaning
behind Eq.~(\ref{eq:Gamma_d}) can be understood as follows. First, a
single channel QPC is in general described by a SU(2) matrix which has
three independent physical variables (just as in any
spin-$1/2$ problem). The third hidden variable $\Delta\varphi$ appears
due to the superposed input. Physically, $\varphi_j$ is the relative phase
between the two amplitudes, $t_j$ and $r_j'$. These are the two amplitudes
injected from the two different inputs and combined into a single output.
Naturally, the sensitivity of this phase appears only by using a
superposed input. In the limit of single input ($p=0$ or $p=1$),
Eq.~(\ref{eq:Gamma_d}) reduces to the existing result
\begin{equation}
 \Gamma_d\rightarrow \Gamma_d^0 = \frac{1}{\Delta t} \left[
     \frac{(\Delta T)^2}{8T(1-T)} + \frac{1}{2}T(1-T)(\Delta\phi)^2
                                                     \right] ,
\end{equation}
where $T=(|t_0|^2+|t_1|^2)/2$.

With the additional phase sensitivity $\Delta\varphi$, we can achieve
an improvement of the overall sensitivity. In the following, we discuss
how it can be done in a systematic way. For simplicity, we consider a low
efficiency limit ($\Delta T\ll \Delta\phi, \Delta\varphi$), where the direct
current measurement through the QPC extracts only a very small portion of the
charge state information. This limit is meaningful because of the great potential
for improvement of detection by controlling the interference. In addition,
it has been argued~\cite{Kang05,Kang07} that a generic
QPC would show a low efficiency, which has also been observed experimentally
with its efficiency below $5\%$~\cite{Chang08}.

In this limit ($\Delta T\rightarrow 0$), $\Gamma_d$ of Eq.~(\ref{eq:Gamma_d}) is reduced to
$\Gamma_d=\{u_2(\Delta\phi)^2 + u_3(\Delta\varphi)^2
+ u_6(\Delta\phi\Delta\varphi)\}/\Delta t$.
This value of $\Gamma_d$ can be controlled
by the two input parameters $p$ and $\theta$ of the input state
(Eq.~(\ref{eq:input})). It is straightforward to find that the
maximum dephasing rate (maximum sensitivity)
\begin{subequations}
\begin{equation}
 \Gamma_d^M = \frac{1}{2\Delta t}   \left\{
   \frac{1}{4} \left[ (\Delta\phi)^2+(\Delta\varphi)^2 \right]
   + (T-1/2)(\Delta\phi)(\Delta\varphi) \right\}
\label{eq:dmax}
\end{equation}
is achieved for
the particular input state
$|\chi_{in}\rangle=|\chi_{in}^M\rangle$:
\begin{equation}
 |\chi_{in}^M\rangle
   = \frac{1}{\sqrt{2}} (c_\alpha^\dagger-ie^{i\varphi_0} c_\beta^\dagger)
     |F\rangle~ .
\label{eq:chi_in_m}
\end{equation}
 Notably, $\Gamma_d^M$ is always
larger than $\Gamma_d^0$ (the dephasing rate of the qubit state when
the conventional input ($p=0$ or $p=1$) is used):
$ \Gamma_d^0 = \frac{1}{2\Delta t} T(1-T)(\Delta\phi)^2$ .
The amount of the sensitivity enhancement is found to be
\begin{equation}
 \Delta \Gamma_d \equiv \Gamma_d^M-\Gamma_d^0
   = \frac{1}{2\Delta t} \left[
      (2T-1)\Delta\phi+\Delta\varphi \right]^2 .
\label{eq:DGamma}
\end{equation}
That is, the sensitivity enhancement
depends on the parameters $\Delta\phi, \Delta\varphi$ and $T$. Interestingly,
a sensitivity enhancement is obtained even for $\Delta\varphi=0$, where the third
variable $\varphi_j$ has no charge sensitivity.
\label{eq:max-sens}
\end{subequations}

Since the variables $\Delta\phi$ and $\Delta\varphi$ can be determined
experimentally, a systematic improvement of the sensitivity is possible.
Later we will briefly discuss how it can be done experimentally.
The relation between $\Delta\phi$ and $\Delta\varphi$ is
not universal but depends on the details of the qubit-QPC interaction. Here we consider a
simple potential shift model~\cite{Kang07} where an extra charge of
a qubit provides a uniform shift of the potential. This model is suitable for describing
the low efficiency limit of ($\Delta T\rightarrow0$) charge detection~\cite{Kang07}.
In this model, one can find that $\Delta\varphi=-\Delta\phi$~\cite{Lee12}.
Fig.~2 displays a plot of the dephasing
rate $\Gamma_d$ as a function of $p$ and $\theta$ for this case ($\Delta\varphi=-\Delta\phi$).
The maximum dephasing rate
$\Gamma_d^M$ is achieved for $p=1/2$ and $\theta = \varphi_0-\pi/2$, that is,
for $|\chi_{in}\rangle=|\chi_{in}^M\rangle$, which is
consistent with Eq.~(\ref{eq:max-sens}).
%

The setup of Fig.~1 is not enough for an actual measurement of the charge state.
It can be overcome by putting a {\em measurement} QPC,
(labeled as QPC$_m$), to compose an
interference between the transmitted and the reflected waves (see Fig.~3).
This scheme is particularly useful in the limit of low efficiency of the
QPC interacting with the qubit. For a conventional input scheme of electrons
($p=0$ or $p=1$ limit in our setup), it has been theoretically shown
in Ref.~\onlinecite{Averin05} that
the full amount of information can be extracted (=``quantum limited detection (QLD)")
by controlling QPC$_m$. In the following, we show that a QLD is also possible
in our scheme, with the improved sensitivity.

With a {\em measurement} QPC (QPC$_m$), the scattering matrix of the
interacting QPC, $S_j$, of Eq.~(\ref{eq:Sj}) is transformed as
\begin{equation}
 S_j \longrightarrow S^m S_j ,
\end{equation}
where
\begin{equation}
{S^m}=
\left(\begin{array}{ccc}
r^m & t'^m \\
t^m & r'^m \\
\end{array} \right)
\label{eq:Sm}
\end{equation}
is the scattering matrix of QPC$_m$.

The most interesting case is to inject the maximally sensitive input state, $|\chi_{in}^M\rangle$,
of Eq.~(\ref{eq:chi_in_m}). For this particular input state,
the probe electron state is transformed to the output
\begin{subequations}
\begin{equation}
  |\bar{\chi}_j\rangle = (\bar{r}_j c_\gamma^\dagger
    +\bar{t}_j c_\delta^\dagger) |F\rangle ,
\end{equation}
where
\begin{eqnarray}
  \bar{r}_j &=& \frac{1}{\sqrt{2}}
    \{ r^m r_j + t'^m t_j -i e^{i\varphi_0}(r^m t'_j + t'^m r'_j) \}, \\
  \bar{t}_j &=& \frac{1}{\sqrt{2}}
    \{ t^m r_j + r'^m t_j - ie^{i\varphi_0}(t^m t'_j + r'^m r'_j) \}.
\end{eqnarray}
\end{subequations}
Note that the dephasing rate of Eq.~(\ref{eq:Gamma_d}) is invariant upon
scattering at QPC$_m$, due to the unitarity of $S^m$. After
passing through QPC$_m$, the output state $|\chi_j\rangle$ (Eq.~(\ref{eq:output}))
is transformed to $S^m|\chi_j\rangle$. However, the scalar product
($\lambda$) of the two detector states (Eq.~(\ref{eq:lambda})) is invariant because
$ \lambda\rightarrow\bar{\lambda}
  = \langle\chi_1| S^{m\dagger} S^m |\chi_0\rangle
  = \langle\chi_1|\chi_0\rangle = \lambda$.

%
The QLD can be achieved from the condition
 $\Delta\bar{\phi} \equiv \bar{\phi}_1 - \bar{\phi}_0 = 0$,
where $\bar{\phi}_j = \arg{\bar{t}_j/\bar{r}_j}$. This is the relation
that the measurement rate reaches the dephasing rate~\cite{Korotkov01,Pilgram,Averin05,Khym06}.
We find that this leads to the condition
\begin{equation}
 \Delta\bar{\phi} = \frac{1-2T^m}{1-4T^m(1-T^m)\sin^2\Theta}\Delta\phi = 0,
\end{equation}
where $T^m=|t^m|^2$ and $\Theta=arg(t^m/r'^m)-arg(t_0/r_0)$.
Therefore, the QLD can be easily achieved by tuning the transmission
probability of the {\em measurement} QPC as
\begin{equation}
 T^m = 1/2.
\label{eq:qld}
\end{equation}
The two conditions,
Eq.(\ref{eq:chi_in_m}) and Eq.~(\ref{eq:qld}), provide
the {\em ultimate sensitivity and efficiency} that can be extracted from
a generic single-channel QPC.

Finally, we briefly describe how this ultimate scheme of maximum sensitivity and efficiency
can be experimentally realized. In practice, we need three quantum point
contacts that form a double interference scheme (see Fig.~3), which is an extension of
the electronic Mach-Zehnder interferometer~\cite{Ji03}.
The superposed input state is generated by QPC$_i$
(``{\em input} QPC"). The maximally sensitive input state $|\chi_{in}^M\rangle$
(Eq.~(\ref{eq:chi_in_m})) can
be easily prepared by controlling QPC$_i$. This input state is interacting
with the qubit at the ``{\em main} QPC". The efficiency is independently
controlled with QPC$_m$, the ``{\em measurement} QPC".

Further, the phase sensitivities $\Delta\phi$ and
$\Delta\varphi$ (or equivalently, $\phi_j$ and $\varphi_j$ with the two
charge states $j=0,1$) can be measured in the setup of Fig.~3 as follows.
$\phi_j=\arg{t_j/r_j}$ is the relative phase between the two split waves (at
the {\em main} QPC) of a single incident wave. This can be directly
achieved by injecting a conventional input state with $p=1$
($|\chi_{in}\rangle=c_\alpha^\dagger|F\rangle$) (or with $p=0$
($|\chi_{in}\rangle=c_\beta^\dagger|F\rangle$)).
The phase $\phi_j$ appears in the interference pattern at the output electrode,
with the condition $0<T_m<1$.
On the other hand, $\varphi_j=\arg{t_j/r_j'}$ corresponds to the relative
phase of
the two {\em merged} waves initially incident from the two separated
inputs $\alpha$ and $\beta$. This phase shift can be extracted by tuning
$0<p<1$ and $T_m=0$ (or $T_m=1$).  This measurement of $\phi_j$ and $\varphi_j$ would allow
a quantitative study of the controlled dephasing and measurement discussed in our proposal.


In conclusion,
we have investigated the ultimate sensitivity and efficiency of a single-channel
QPC as a charge detector. In contrast to the conventional
charge detection schemes that utilize only one or two variables, we have
shown that a QPC provides three independent physical variables for charge
detection, due to the SU(2) symmetry of a scattering matrix. The hidden third
information is revealed by injecting a superposed input state of the probe electrons.

This work was supported by the
National Research Foundation of Korea under Grant 
No.~2009-0084606, 2012R1A1A2003957, and by LG Yeonam Foundation.

\begin{figure}[b]
\includegraphics[width=3.5in]{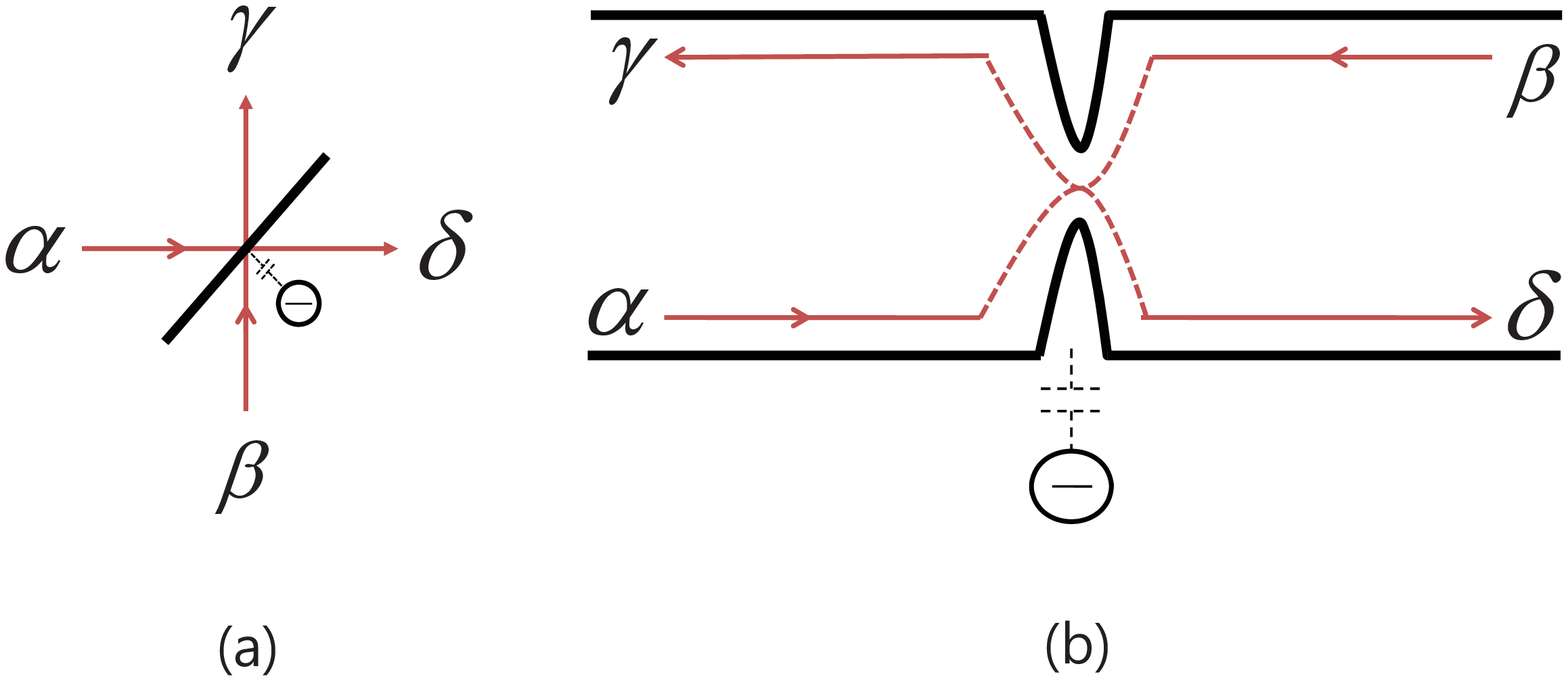}
\caption{\label{fig1}
 (a) Charge sensing scheme of a quantum point contact with a superposed input
 state, and
(b) a possible realization with the quantum Hall edge state
 (Color online).}
\end{figure}

\begin{figure}[b]
\includegraphics[width=3.5in]{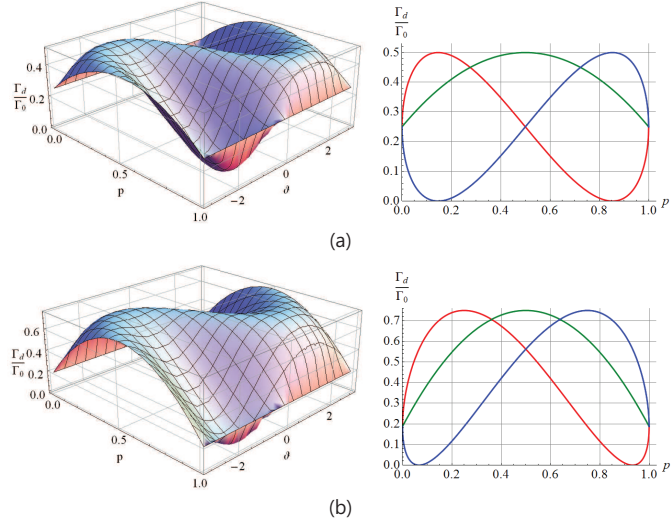}
\caption{\label{fig2}
 Dephasing rate ($\Gamma_d$) with the potential shift
 model ($\Delta\varphi=-\Delta\phi$) as a function of the two input parameters,
$p$ and $\vartheta~(\vartheta\equiv\varphi_0-\theta-\pi/2)$ for (a) $T={1\over2}$, and for (b)  $T={1\over4}$.
3D plots of the dephasing rate $\Gamma_d$ (in unit of $\Gamma_0\equiv (\Delta\varphi)^2/(2\Delta t)$)
are given in the left panels.
The right panels of (a) and (b) display the dephasing rate as a function of $p$ for three different values of the input phase $\vartheta=-\pi/2$ (red), $0$ (green), $\pi/2$ (blue), respectively
(Color online).
}
\end{figure}

\begin{figure}[b]
\includegraphics[width=3in]{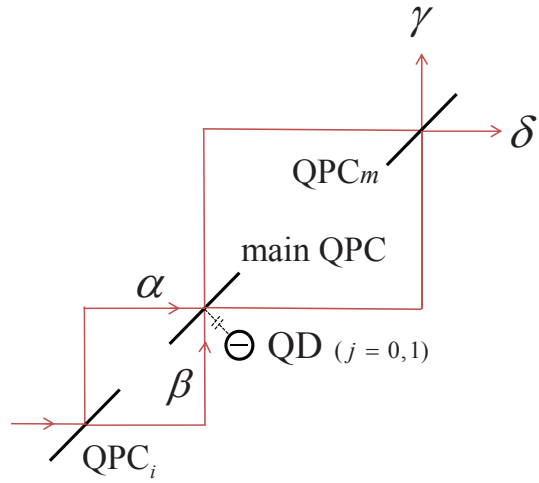}
\caption{\label{fig3}
Schematic of a charge detection setup with full control of the sensitivity and the efficiency.
The setup consists of the three QPCs, namely the {\em input} QPC (QPC$_i$),
the main QPC, and the {\em measurement} QPC (QPC$_m$) (Color online).
}
\end{figure}


\begin{thebibliography}{99}
\bibitem{Delta_t} This time scale is based on a spinless model. If the spin
degeneracy is taken into account, it becomes $\Delta t = h/(2eV)$.
\end{thebibliography}
\end{document}